\documentclass[a4paper, 11pt,onecolumn,final,notitlepage,oneside]{article}
\usepackage[cp1251]{inputenc}
\usepackage[T1,T2A]{fontenc}
\usepackage[english]{babel}
\usepackage{indentfirst}
\usepackage{geometry}
\usepackage{graphicx}
\usepackage[final]{epsfig}
\usepackage{multicol}
\usepackage{graphicx}
\usepackage{subfigure}
\usepackage{amsmath}
\usepackage{enumerate}
\usepackage{amssymb}
\usepackage{amscd}
\usepackage{hhline}
\usepackage{multirow}
\usepackage{setspace}
\usepackage{makeidx}
\usepackage{textcomp}
\usepackage{amsfonts}
\usepackage{afterpage}
\usepackage{longtable}
\usepackage{cite}
\usepackage{rawfonts}
\usepackage{array}
\usepackage{amsopn}
\usepackage{hyperref}
\hypersetup{colorlinks=true, linkcolor=blue, citecolor=green, filecolor=magenta, urlcolor=magenta, pdfpagemode=FullScreen}

 \graphicspath{{pictures/}}

\DeclareGraphicsExtensions{.pdf,.png,.jpg,.eps}

\numberwithin{equation}{subsubsection}

\makeatletter
\def\@seccntformat#1{\csname the#1\endcsname.\ } 
\def\@biblabel#1{#1.} 
\makeatother

\begin{document}

\title{\normalsize \begin{flushleft}
{UDC 534.014.1, 534.5\\PACS 03.30.+p}
\end{flushleft}
\vspace{\baselineskip}
 \normalsize \bf Some ways parameter calculation curvilinear uniformly accelerated motion}
\author{\bf \small V.\,V.~Voytik\,\\
\small \itshape Departement of Medical Physics and Informatics,\\ \small \itshape Bashkir State Medical University, \\ \small \itshape Lenin st. 3, Ufa, 450008, Russian Federation\\
\small \itshape e-mail: voytik1@yandex.ru, vvvojtik@bashgmu.ru\\
\small \itshape Received 29.06.2015 г.}
\date{}
\maketitle
\renewcommand{\abstractname}{}

\begin{abstract}
The parameters of the general curvilinear motion in the laboratory frame $S^*$ of uniformly accelerated reference frame $s$  three equivalent ways is calculated. One of these methods is a boost from the frame reference in which $s$ is moving in a straight line while maintaining proper orientation. Another method is to solve the equations of inverse problem kinematics. The third method is to solve the equations of motion of the tetrad. This shows that these equations can be considered reasonable. The article also found explicitly transformation to a curvilinear moving uniformly accelerated reference frame and proved the assertion that in the frame reference $S^*$,  Thomas precession and Wigner rotation  $s$ in opposite directions and cancel each other out.
\end{abstract}

\begin{flushleft}
{\bf{Introduction}}
\end{flushleft}

\textit{Uniformly accelerated} motion, as is known, is a motion of a rigid frame of reference in which the components of the proper acceleration of its origin remain constant. Moreover, such a frame of reference does not rotate relative to an instantaneously accompanying inertial frame of reference. Sometimes \cite[p. 128, par. 19]{1} such motion is understood to mean motion under the action of a force that remains constant in the laboratory frame. However, if the frame of reference under the action of the force does not move in a straight line, and since during the motion the direction of the velocity of the frame of reference changes relative to the axes of the laboratory frame $S^*$, then, due to the tensor law of transformation of 3-forces, the force in the proper frame of reference will change both in magnitude and in direction. Therefore, in general, such motion cannot be considered as uniformly accelerated. This motion - under the action of a uniform force field in the laboratory frame of reference will here be called \textit{hyperbolic}. The difference between these two types of motion of the reference frame $s$ is also evident from the fact that for hyperbolic motion in the $S^*$ system, the proper orientation of $s$ is usually irrelevant, and is therefore completely determined by the initial velocity $s$. Uniformly accelerated motion $s$ is characterized, in addition to the initial velocity, by three angles that determine the initial orientation of $s$ relative to $S^*$. Thus, a distinction must be made between uniformly accelerated and hyperbolic motion. Only when hyperbolic motion and uniformly accelerated motion of different reference frames occur along the same straight line will the velocities of their reference frames as functions of laboratory time coincide with each other.\footnote[2]{This statement will be proven in another article.}

The goals of this article are twofold.

Previously, in \cite[formulas (2.10), (2.11)]{2}, the so-called "inverse kinematics equations" were presented, that is, differential equations for the parameters of motion of a non-inertial reference frame: the 3-vector $\mathbf{v'}$ and the rotation matrix $a_{\alpha\beta}$. On the other hand, the equation of motion for an orthonormal tetrad of 4-vectors associated with a given reference frame \cite[formula (4)]{12} has long been known (see also \cite[formulas (5.3), (5.4)]{21}). All these equations are consequences (see \cite{2}, \cite[formulas (15), (16)]{13}) of the well-known Lorentz-M{\o}ller-Nelson transformation into an arbitrary rigid frame of reference and, therefore, are equivalent to each other. However, these equations have hardly been used in practice (see, however, \cite{23}, \cite{20}) and, therefore, they are still insufficiently substantiated. Therefore, one of the goals of the article is to test these equations using the example of the most general uniformly accelerated motion. The reason for using uniformly accelerated motion as a test case is the mathematical difficulties in solving general differential equations of the inverse kinematics problem.

Moreover, general (not necessarily rectilinear) uniformly accelerated motion is interesting in itself. In practice, the motion of a uniformly accelerated reference frame may be encountered in future astronautics, since space stations, when flying, will be systems with constant proper acceleration. The problem of uniformly accelerated motion is used, for example, in the study of the electromagnetic radiation of a charge. Due to its importance, uniformly accelerated motion was considered in the earliest stages of the development of the theory of relativity. The pioneering work in which the correct relativistic transformation into a rectilinearly moving uniformly accelerated reference frame was obtained is \cite{25}. Subsequently, based on various initial assumptions, many attempts were made to consider the general transformation into a uniformly accelerated reference frame. However, it should be recognized that, for some reason, the general transformation into a uniformly accelerated reference frame is not explicitly written down. Therefore, the second goal of the article is to obtain, without any simplifications or assumptions, an explicit transformation into a uniformly accelerated reference frame $s$ moving curvilinearly relative to the laboratory frame of reference $S^*$.

This article is structured as follows.

The first section of this article provides preliminary information regarding the behavior of the Lorentz-M{\o}ller-Nelson transformation parameters during a boost. In Section 2, the general uniformly accelerated motion and all its parameters are obtained from the initial rectilinear uniformly accelerated motion using a simple Lorentz boost. Due to their clarity, the parameters of the general uniformly accelerated motion obtained using the boost will serve as a model for comparison with two other methods for determining uniformly accelerated motion. One of these methods involves directly solving the inverse kinematics equations in Section 4, and the second (Section 5) involves solving the equations of motion for a tetrad. A comparison of the solutions obtained by all these methods is performed in Section 6. Finally, Section 7 presents the general transformation to a uniformly accelerated rigid reference frame. In addition, in the third paragraph one statement of the article \cite{21} about the mutual compensation of the Wigner rotation and the Thomas precession for the uniformly accelerated motion of the system $s$ will be verified.

\subsubsection{Change in the special Lorenz-M{\o}ller-Nelson transformation when have boost}

Let us recall how the so-called special Lorentz-M{\o}ller-Nelson transformation (hereinafter $c=1$) changes during a boost from the laboratory frame of reference $S:(T,\mathbf{R})$ to a non-inertial, radially rigid frame of reference $s:(t,\mathbf{r})$, which moves arbitrarily “without rotation” relative to $S$ \cite{33}
\begin{equation} \label{1}
     T=\frac{\mathbf{vr}}{\sqrt{1-v^{2} } } +\int _{0}^{t}\frac{dt}{\sqrt{1-v^{2} } }\,\,,
\end{equation}
\begin{equation} \label{2}
 \mathbf{R=r}+\frac{1-\sqrt{1-v^{2} } }{v^{2} \sqrt{1-v^{2} } } \mathbf{(vr)v}+\int _{0}^{t}\frac{\mathbf{v}dt}{\sqrt{1-v^{2} } }\,\, .
\end{equation}
Let us make the transition from the reference frame $S$ to the inertial reference frame $S^*:(T^*,\mathbf{R^*})$ moving with a constant velocity $\mathbf{u}$ relative to it. The coordinates and time in the reference frames $S$, $S^*$ are related by the usual Lorentz transformation
\begin{equation}\label{6}
  T^*=\frac{T-\mathbf{uR}}{\sqrt{1-u^2}}\,\,,
\end{equation}
\begin{equation}\label{7}
  \mathbf{R^*=R}+\frac{1-\sqrt{1-u^2}}{u^2\sqrt{1-u^2}}\,\mathbf{(uR)u}-\frac{\mathbf{u}T}{\sqrt{1-u^2}}\,\,.
\end{equation}
It turns out that the reference systems $S^*$ and $s$ are related by a common Lorentz-M{\o}ller-Nelson transformation of the form
\begin{equation} \label{4}
T^*=\frac{v_{\alpha }^* a_{\beta \alpha} r'_{\beta } }{\sqrt{1-v^{*2} } } +\int _{0}^{t}\frac{dt}{\sqrt{1-v^{*2} } }\,\,,
\end{equation}
\begin{equation} \label{5}
 R_{\alpha }^* =a_{\beta \alpha} r'_{\beta } +\frac{1-\sqrt{1-v^{*2}}}{v^{*2} \sqrt{1-v^{*2}}} \,v_{\alpha }^* v_{\gamma }^* a_{\beta \gamma} r'_{\beta} +\int _{0}^{t}\frac{v_{\alpha }^* dt}{\sqrt{1-v^{*2}}}\,\,,
  \end{equation}
where
\begin{equation} \label{13}
\mathbf{v^*}=\frac{\sqrt{1-u^{2} }\mathbf{v-u}}{1-\mathbf{uv}} +\frac{(1-\sqrt{1-u^{2} } )\mathbf{(uv)u}}{u^{2} (1-\mathbf{uv})}\,\,.
\end{equation}
The meaning of $\mathbf{v^*}$ is that this quantity is the velocity of the origin of reference frame $s$ in the new laboratory frame $S^*$ expressed in terms of proper time. The matrix $a_{\alpha\beta}$ is the rotation matrix of frame $s$ relative to the new frame $k$, the transformation to which is special (i.e., of the form \eqref{1}, \eqref{2}) with parameter $\mathbf{v^*}$. In this case, the origins of frames $s$ and $k$ coincide and it is conventionally assumed that the axes of frame $k$ are oriented without rotation relative to $S^*$. The quantity $v'^*_{\beta}=v_{\alpha }^* a_{\beta \alpha}$ is the velocity vector of the origin of reference frame $s$ defined in coordinate system $s$ and
\begin{equation} \label{14}
\mathbf{v'^*}=\frac{\mathbf{v}-\sqrt{1-v^{2} } \mathbf{u}}{1-\mathbf{uv}} -\frac{(1-\sqrt{1-v^{2} })\mathbf{(uv)v}}{v^{2}(1-\mathbf{uv})}\,\, .
\end{equation}
The rotation matrix $a^{\beta\alpha}$ in coordinates: rotation axis $\mathbf{n}$ , rotation angle $\phi$ has the form
\begin{equation} \label{2.1}
a^{\beta\alpha}=\delta^{\alpha\beta}\cos\phi_W+n^{\beta}n^{\alpha}(1-\cos\phi_W)-e^{\alpha\beta\gamma}n^{\gamma}\sin\phi_W\,.
  \end{equation}
The rotation angle, which is called the Wigner rotation and corresponds to the matrix $a_{\beta\alpha}$ is equal to
\begin{equation} \label{31}
\mathbf{n}\sin \phi_W=\frac{\mathbf{u\times v}}{1+\sqrt{1-v^{2}}\sqrt{1-u^2}-\mathbf{uv}}\cdot
\left(1-\frac{(1-\sqrt{1-u^2})(1-\sqrt{1-v^{2}})}{u^2v^{2}}\mathbf{\,\,uv}\right),
  \end{equation}
\begin{equation} \label{32}
 \cos \phi_W=\frac{\sqrt{1-u^2}+\sqrt{1-v^{2}}-\mathbf{uv}+\frac{(1-\sqrt{1-u^2})(1-\sqrt{1-v^{2}})}{u^2v^{2}}\mathbf{(uv)}^2} {1+\sqrt{1-u^2}\sqrt{1-v^{2}}-\mathbf{uv}}\,\,,
 \end{equation}
\begin{equation} \label{33}
\mathbf{n}\,\tan \frac{\phi_W }{2}=\mathbf{n}\frac{\sin \phi_W }{1+\cos \phi_W }=\frac{\mathbf{u}\times \mathbf{v}}{\left( 1+\sqrt{1-{{u}^{2}}} \right)\left( 1+\sqrt{1-{{v}^{2}}} \right)-\mathbf{uv}}\,\,,
\end{equation}
where $\mathbf{n}$ is the unit vector in the direction of rotation.

\subsubsection{Determining the parameters of uniformly accelerated motion using boost}

Let us formulate the problem as follows. Let us first choose the proper orientation of the uniformly accelerated reference frame $s:(t,\mathbf{r})$ so that its proper acceleration is directed along the 1-axis. Let the uniformly accelerated reference frame $s:(t,\mathbf{r})$ initially be at rest in some laboratory reference frame $S:(T,\mathbf{R})$ and completely coincide with $S$. Then it moves rectilinearly relative to $S$ with proper acceleration $W$ in the positive direction of the 1-axis. In this case, the uniformly accelerated motion is well known. Its initial velocity is equal to $v_1=\mathrm{th} \,Wt$ \cite[p. 207, formula (8.168)]{7}. Let us now transition from frame $S$ to a new inertial reference frame $S^*:(T^*,\mathbf{R^*})$, which moves in the plane of axes 1 and 3 with velocity $\mathbf{u}=(u_1,0,u_3)$. With this boost, frame $s$ in the new reference frame acquires initial velocity components along axes 1 and 3. Consequently, the motion of frame $s$ in frame $S^*$, which can be considered the laboratory frame, will be a general uniformly accelerated motion. It is only necessary to determine the new parameters of $s$'s motion in $S^*$.

Now knowing formulas \eqref{13}-\eqref{33}, it is easy to find the parameters of curvilinear uniformly accelerated motion. According to the formula \eqref{13} the new velocity $s$ relative to $S^*$ will be equal to (we will write the Greek indices of the vector components below to distinguish them from the exponents)
\begin{equation} \label{36}
v^*_{1}=\frac{\sqrt{1-u^2}\,\,\mathrm{th}Wt-u_1}{1-u_1\mathrm{th}Wt}+\frac{\left( 1-\sqrt{1-{{u}^{2}}} \right)u_{1}^{2}\,\mathrm{th}Wt}{u^2(1-u_1\mathrm{th}Wt)}\,\,,
\end{equation}
\begin{equation} \label{37}
v^*_{2}=0\,\,,
\end{equation}
\begin{equation} \label{38}
v^*_{3}= -\frac{u_3}{1-u_1\mathrm{th}Wt}+\frac{\left(1-\sqrt{1-u^2}\right)u_1u_3 \,\mathrm{th}Wt}{u^2(1-u_1\mathrm{th}Wt)}\,\,.
\end{equation}
The same speed expressed in the $s$ coordinate system is equal according to \eqref{14}
\begin{equation} \label{39}
v'^*_1=\frac{v_1-u_1}{1-u_1v_1}=\frac{\mathrm{th}Wt-u_1}{1-u_1\,\mathrm{th}Wt}\,\,,
\end{equation}
\begin{equation} \label{40}
v'^*_2=0\,\,,
\end{equation}
\begin{equation} \label{41}
v'^*_3=-\frac{\sqrt{1-v^2}\,\,u_3}{1-u_1v_1}=-\frac{u_3}{\mathrm{ch}Wt-u_1\,\mathrm{sh}Wt}\,\,.
\end{equation}
At the same time
\begin{equation} \label{41.3}
v_{\alpha}=a_{\beta\alpha}v'_\beta\,\,,
\end{equation}
and the Wigner rotation matrix of the system $s$ relative to $S^*$ is equal to
\begin{equation} \label{44.1}
	  a_{\beta\alpha} = \begin{pmatrix}
	\cos\phi_W&0&-\sin\phi_W \\
	0      & 1 &    0       \\
		\sin\phi_W & 0 & \cos\phi_W \\
	\end{pmatrix}
\end{equation}
The Wigner rotation that must be performed during the boost between the coordinate system $s$ and the coordinate system that moves "without rotation" relative to $S$ is carried out in the plane of axes 1 and 3 in the direction $3\rightarrow 1$ . According to \eqref{31}-\eqref{33}, the angle of this rotation satisfies the equalities
\begin{equation} \label{42}
\sin {{\phi_W }}=\frac{{{u}_{3}}\left[\,\,\mathrm{sh}Wt-\frac{1-\sqrt{1-{{u}^{2}}}}{{{u}^{2}}}\,\,{{u}_{1}}\left( \mathrm{ch}Wt-1 \right)\,\right]}{\mathrm{ch}Wt-{{u}_{1}}\,\mathrm{sh}Wt+\sqrt{1-{{u}^{2}}}}\,\,,
\end{equation}
\begin{equation} \label{43}
\cos {{\phi_W }}=\frac{\left( \sqrt{1-{{u}^{2}}}+\frac{1-\sqrt{1-{{u}^{2}}}}{{{u}^{2}}}\,u_{1}^{2} \right)\mathrm{ch}Wt-{{u}_{1}}\,\mathrm{sh}Wt+1-\frac{1-\sqrt{1-{{u}^{2}}}}{{{u}^{2}}}\,\,u_{1}^{2}}{\mathrm{ch}Wt-{{u}_{1}}\,\mathrm{sh}Wt+\sqrt{1-{{u}^{2}}}}\,,
\end{equation}
\begin{equation} \label{44}
\mathrm{tg}\frac{\phi_W}{2}=\frac{{{u}_{3}}\,\,\mathrm{sh}Wt}{\left( 1+\sqrt{1-{{u}^{2}}} \right)\left(1+ \mathrm{ch}Wt\right)-{{u}_{1}}\,\mathrm{sh}Wt}\,\,.
\end{equation}

From this it is clear that the initial orientation angle of the uniformly accelerated system obtained by boosting from \eqref{44} is zero.
By calculating the vectors $\mathbf{v^*}$ and the Wigner angle $\phi_W$, the uniformly accelerated motion is completely specified. The parameters of the uniformly accelerated motion obtained in this way can serve as a model for comparison with other solutions.

\subsubsection{Mutual compensation of Thomas precession and Wigner rotation in the case of uniformly accelerated motion}

Since uniformly accelerated motion does not have its own rotation, the Wigner rotation and Thomas precession must compensate each other. We will demonstrate this by direct calculation. The angular velocity of the Wigner rotation can be obtained by differentiating equation \eqref{42}. We obtain
\[{{\omega_W }}\cos {{\phi_W }}=\frac{W{{u}_{3}}\left[ \left( \sqrt{1-{{u}^{2}}}+\frac{1-\sqrt{1-{{u}^{2}}}}{{{u}^{2}}}u_{1}^{2} \right)\mathrm{ch}Wt-{{u}_{1}}\mathrm{sh}Wt+1-\frac{1-\sqrt{1-{{u}^{2}}}}{{{u}^{2}}}u_{1}^{2} \right]}{{{\left[ \mathrm{ch}Wt-{{u}_{1}}\mathrm{sh}Wt+\sqrt{1-{{u}^{2}}}\, \right]}^{2^{}}}}\,\,.\]

Dividing this expression by \eqref{43} we get
\begin{equation} \label{46}
{{\omega }}_W=\frac{W{{u}_{3}}}{\mathrm{ch}Wt-{{u}_{1}}\mathrm{sh}Wt+\sqrt{1-{{u}^{2}}}}\,\,.
\end{equation}
Let us now calculate the angular velocity of Thomas precession $\mathbf{\Omega}_T$
\begin{equation} \label{46.12}
\mathbf{\Omega}_T=\frac{1-\sqrt{1-v^{*2}}}{v^{*2}\sqrt{1-v^{*2}}}\,\,\mathbf{v}^*\times\mathbf{\dot{v}}^*\,.
\end{equation}
The first factor in its definition will be equal to
\begin{equation} \label{47}
\frac{1-\sqrt{1-{{v}^{*2}}}}{{{v}^{*2}}\sqrt{1-{{v}^{*2}}}}=\frac{1}{\sqrt{1-{{u}^{2}}}}\cdot\frac{{{\left[\,\mathrm{ch}Wt-{{u}_{1}}\mathrm{sh}Wt \right]}^{\,2}}}{\mathrm{ch}Wt-{{u}_{1}}\mathrm{sh}Wt+\sqrt{1-{{u}^{2}}}}\,\,.
\end{equation}
Differentiating the speed parameter we obtain
\begin{equation} \label{48}
\dot{v}^*_{1}=\frac{W\sqrt{1-{{u}^{2}}}}{{{\left(\mathrm{ch}Wt-{{u}_{1}}\mathrm{sh}Wt \right)}^{2}}}\left[ 1-\frac{\left( 1-\sqrt{1-{{u}^{2}}} \right)u_{1}^{2}}{{{u}^{2}}} \right],
\end{equation}
\begin{equation} \label{49}
\dot{v}^*_{2}=0\,\,,
\end{equation}
\begin{equation} \label{50}
\dot{v}^*_3=-\frac{\sqrt{1-u^2}\left(1-\sqrt{1-u^2}\right)}{u^2}\cdot \frac{Wu_1u_3}{\left( \mathrm{ch}Wt-u_1\mathrm{sh}Wt\right)^2}\,\,.
\end{equation}
The second factor is a vector product and if we take into account \eqref{36}- \eqref{38}, \eqref{48}- \eqref{50} it will be equal to
\begin{equation} \label{51}
(\mathbf{v}^*\times \mathbf{\dot{v}^*})_2=v^*_{3}\dot{v}^*_{1}-v^*_{1}\dot{v}^*_{3}=-\frac{W\sqrt{1-u^2}\,u_3}{\left(\mathrm{ch}Wt-u_1\mathrm{sh}Wt\right)^2}\,\,.
\end{equation}
Therefore, multiplying \eqref{47} and \eqref{51} we obtain that the Thomas precession is equal to
\begin{equation} \label{52}
{{ {{\Omega }_{T}} }}=-\,\frac{W{{u}_{3}}}{\,\mathrm{ch}Wt-{{u}_{1}}\mathrm{sh}Wt+\sqrt{1-{{u}^{2}}}}\,\,.
\end{equation}
Comparing formulas \eqref{46} and \eqref{52} it is clear that the sum of the proper Thomas precession and the proper Wigner rotation of the system $s$ is zero, as it should be.

\subsubsection{Determination of motion parameters by solving the equations of the inverse kinematics problem}

Let us now consider the application of differential equations of the inverse kinematics problem \cite{2} to curvilinear uniformly accelerated motion
\begin{equation}\label{66.1}
		\mathbf{W'}=\mathrm{const}\,\,\,,\,\,\mathbf{\Omega'}=0\,\,. 
		\end{equation}
Taking this condition into account, the differential equations of the inverse kinematics problem \cite[formulas (2.10), (2.11)]{2} take the form
\begin{equation}\label{66}
       \mathbf{\dot{v}'}=\frac{d\mathbf{v'}}{dt}=\mathbf{W'-(W'v')v'}\,\,,
    \end{equation}
    \begin{equation}\label{67}
      \boldsymbol{\omega}'=-\frac{1-\sqrt{1-v'^2}}{v'^2}\,\mathbf{v'\times W'}\,\,.
    \end{equation}
According to par. 2, we choose axis 1 of the laboratory reference frame $S$ along the acceleration vector $\mathbf{W'}$ of the system $s$, and axis 3 – along the trajectory lying in its plane. Equation \eqref{66} for axis 1 is
\[\dot{v}'_1=W(1-v'^2_1)\,\,.\]
Integrating this equation we get
  \begin{equation} \label{68}
 {{v'}_{1}}(t)=\ \mathrm{th}(Wt+k )\,\,,
 \end{equation}
where
 \[ \mathrm{th} k ={{v'}_{1}}(0)\,\,.\]
Projecting equation \eqref{66} onto axis 3 perpendicular to acceleration, we obtain, taking into account \eqref{68}, that
\[{{\dot{v'}}_{3}}+W\ \mathrm{th}\,(Wt+k )\,{{v'}_{3}}=0\,\,.\]
From here
 \begin{equation} \label{69}
    {{v'}_{3}}(t)={{v'}_{3}}(0)\frac{\mathrm{ch}\,k }{\mathrm{ch}\,(Wt+k )}\,\,,
     \end{equation}
 \[v'^2=v'^2_1+v'^2_3=\frac{v'^{\,2}_3(0)\mathrm{ch}^2k +\mathrm{sh}^2(Wt+k)}{\mathrm{ch}^2(Wt+k)}\,\,,
\sqrt{1-{{{{v}'}}^{2}}}=\frac{\sqrt{1-{{{{v}'}}^{2}}(0)}\ \mathrm{ch} k}{\mathrm{ch}(Wt+k)}\,\,. \]
Substituting these values into the expression \eqref{67} we obtain that the only non-zero component of the vector $\boldsymbol{\omega}'_W$ is
\begin{equation} \label{70}
 {\omega' }_{2W}=\dot{\phi}_W=-\frac{1-\sqrt{1-{{{{v}'}}^{2}}}}{{{{{v}'}}^{2}}}\,{v'}_3 {W}=-\frac{Wv'_3(0)\, \mathrm{ch} k \ }{\mathrm{ch} (Wt+k )+\sqrt{1-{v'}_{3}^{2}(0)\ {\mathrm{ch}^{2}}k } }\,\,.
  \end{equation}
Integrating this expression we get that
 \begin{equation} \label{71}
   \phi_W =2\,\mathrm{arctg} \,\frac{e^{Wt+k }+\sqrt{1-v'^2_3(0) \mathrm{ch} k}}{v'_3(0)\mathrm{ch}^2 k}-2\,\mathrm{arctg}\,\frac{e^k+\sqrt{1-v'^2_3(0)\mathrm{ch} k }}{v'_3(0)\mathrm{ch}^2k }+\alpha \,\,.
      \end{equation}

\subsubsection{Determination of motion parameters by solving the equations of motion for tetrad of 4-vectors}

Uniformly accelerated motion can also be determined immediately in 4-dimensional form by solving the equations of motion of a tetrad of unit 4-vectors associated with the origin of the uniformly accelerated system $s$ \cite[formula (4)]{12}, \cite[formulas (5.3), (5.4)]{21}
\begin{equation}\label{31.1}
  \frac{d\Lambda^{0 i}}{dt}=W^{\alpha}\Lambda^{\alpha i}\,,
\end{equation}
\begin{equation}\label{30.1}
  \frac{d\Lambda^{\alpha i}}{dt}=W^{\alpha}\Lambda^{0i}+e^{\alpha\beta\gamma}\Omega^{\gamma}\Lambda^{\beta i}\,\,,
\end{equation}
under orthonormality conditions
\begin{equation} \label{18.1}
	{{\Lambda }^{0i}}\Lambda ^{0}_{\ \ i}=1,\,\,\,{{\Lambda }^{0i}}\Lambda ^{\alpha }_{\ \ i}=0,\,\,\,{{\Lambda }^{\alpha i}}\Lambda ^{\beta }_{\ \ i}={-\,{\delta }^{\alpha \beta }}\,.
\end{equation}

Four 4-vectors have 16 components, which appear in 16 equations \eqref{31.1}, \eqref{30.1}. These components are related to each other by 1 + 3 + 6 = 10 conditions \eqref{18.1}. Thus, there are a total of 16 - 10 = 6 independent quantities for which 16 - 10 = 6 independent equations can be written, such that these equations admit a solution.
Although in general, equations \eqref{31.1}, \eqref{30.1} are no simpler than the inverse kinematics equations, for a uniformly accelerated frame of reference they are significantly simplified and admit a quick solution.
The idea of using these equations belongs to Friedman and Scarr \cite{23}, \cite{20}.

Taking into account \eqref{66.1}, we differentiate \eqref{31.1} with respect to $t$ and substitute the value of \eqref{30.1} into the resulting expression. We obtain that
\begin{equation}\label{31.2}
  \frac{d^{\,2}\Lambda^{0 i}}{dt^{\,2}}=W^2\Lambda^{0 i}\,,
\end{equation}
The solution to this equation, taking into account the first of the equations \eqref{18.1}, is
\begin{equation}\label{31.3}
  \Lambda^{0 i}=A^{i}\,\mathrm{sh}Wt+B^{i}\,\mathrm{ch}Wt\,,
\end{equation}
\begin{equation} \label{18.2}
	{{A }^{i}}A_{i}=-1,\,\,\,{A^{i}}B_{i}=0,\,\,\,{B^{ i}}B_{i}=1\,.
\end{equation}

Next, we choose a uniformly accelerated coordinate system according to point 2 so that the proper acceleration is directed along the 1-axis. Now, taking into account \eqref{31.1}, we differentiate the solution \eqref{31.3} with respect to $t$. We obtain that
the first 4-vector is equal to
\begin{equation}\label{30.5}
  \Lambda^{1 i}=B^i\,\mathrm{sh}Wt+A^i\,\mathrm{ch}Wt\,.
\end{equation}
The equations for the remaining 4-vectors are trivial
\begin{equation}\label{30.6}
  \frac{d\Lambda^{2 i}}{dt}=\frac{d\Lambda^{3 i}}{dt}=0\,\,.
\end{equation}
The constant 4-vectors $A^{i}$, $B^{i}$, $\Lambda^{2 i}$, $\Lambda^{3 i}$ are determined from the condition that the 4-vectors $\Lambda^{0i}$ , $\Lambda^{\alpha i}$ as functions of time $t$ must have the following form \cite[formulas (2.1), (2.2)]{21} (we write the indices below)
\begin{equation}\label{16}
  \Lambda_{0i}=(\Lambda_{00},\Lambda_{0\alpha})=\left(\frac{1}{\sqrt{1-v^{*2}}}\,,\frac{v^*_{\alpha}}{\sqrt{1-v^{*2}}}\right),
\end{equation}
\begin{equation}\label{17}
  \Lambda_{\alpha i}=(\Lambda_{\alpha 0},\Lambda_{\alpha\beta})=\left(\frac{v^*_{\gamma}a_{\alpha\gamma}}{\sqrt{1-v^{*2}}}\,,\,\,a_{\alpha\beta}+\frac{1-\sqrt{1-v^{*2}}}{v^{*2}\sqrt{1-v^{*2}}}\,\,v^*_{\beta}v^*_{\mu}a_{\alpha\mu}\right)\,.
\end{equation}
Taking into account these equalities and the fact that at the initial moment of time $v_1=-u_1$, $v_3=-u_3$, and the angle of rotation of the uniformly accelerated system relative to the laboratory frame of reference is zero, it is easy to see that
\begin{equation}\label{161}
  A^{i}=\left(-\frac{u_1}{\sqrt{1-u^2}}, 1+\frac{1-\sqrt{1-u^2}}{u^2\sqrt{1-u^2}}\,u^2_1, 0, \frac{1-\sqrt{1-u^2}}{u^2\sqrt{1-u^2}}\, u_1u_3\right)\,,
\end{equation}
\begin{equation}\label{171}
   B^{i}=\left(\frac{1}{\sqrt{1-u^2}}, -\frac{u_1}{\sqrt{1-u^2}}\,, 0, -\frac{u_3}{\sqrt{1-u^2}}\,\right)\,,
\end{equation}
		\begin{equation}\label{161.1}
  \Lambda^{2 i}=\left(0, 0, 1, 0\right)\,,
\end{equation}
\begin{equation}\label{171.1}
  \Lambda^{3 i}=\left(-\frac{u_3}{\sqrt{1-u^2}}, \frac{1-\sqrt{1-u^2}}{u^2\sqrt{1-u^2}}\,u_1u_3, 0, 1+\frac{1-\sqrt{1-u^2}}{u^2\sqrt{1-u^2}}\,u^2_3  \,\right)\,.
\end{equation}

\subsubsection{Equivalence of these methods for determining uniformly accelerated motion}

Let us now compare solutions \eqref{68}, \eqref{69} with the already known solutions \eqref{39}, \eqref{41}. It is easy to see that these solutions coincide (as they should) with the following choice of constants
 \begin{equation} \label{72}
       k =-\,\mathrm{arth}\,{{u}_{1}}\,\,,
       \end{equation}
       \begin{equation} \label{73}
       {{{v}'}_{3}}(0)=-\,{{u}_{3}}\,\,.
       \end{equation}
One can also verify the identity of \eqref{70} and \eqref{46} under the conditions \eqref{72}, \eqref{73}. Consequently, integrating \eqref{70} leads, up to the constant $\alpha$ in \eqref{71}, which determines the initial orientation of the system $s$ relative to $S^*$, to the Wigner angle \eqref{42}-\eqref{44}.
Thus, the two solutions defining uniformly accelerated motion: using a boost and solving the equations of the inverse kinematics problem, are completely identical.

We will show that the methods for determining uniformly accelerated motion using a boost and solving the equations of motion of a tetrad are equivalent. To do this, we first calculate those components of the 4-velocity of the origin \eqref{16} and the 4-vectors of unit unit vectors \eqref{17}, the calculation of which presents no difficulty. Substituting \eqref{36}-\eqref{38} into \eqref{16} and \eqref{17} we obtain that
\begin{equation}\label{74}
  \Lambda_{00}=\frac{1}{\sqrt{1-v^{*2}}}=\frac{\mathrm{ch}Wt-u_1\mathrm{sh}Wt}{\sqrt{1-u^2}}\,\,,
\end{equation}	
				\begin{equation}\label{75}
  \Lambda_{01}=\frac{v^*_{1}}{\sqrt{1-v^{*2}}}=\left(1+\frac{1-\sqrt{1-u^2}}{u^2\sqrt{1-u^2}}\,u^2_1\right)\mathrm{sh}Wt-\frac{u_1}{\sqrt{1-u^2}}\,\mathrm{ch}Wt\,\,,
\end{equation}
		\begin{equation}\label{76}
  \Lambda_{02}=\Lambda_{20}=\frac{v^*_{2}}{\sqrt{1-v^{*2}}}=0\,\,,
\end{equation}
		\begin{equation}\label{77}
  \Lambda_{03}=\frac{v^*_{3}}{\sqrt{1-v^{*2}}}=\frac{1-\sqrt{1-u^2}}{u^2\sqrt{1-u^2}}\cdot u_1u_3\,\mathrm{sh}Wt-\frac{u_3}{\sqrt{1-u^2}}\,\mathrm{ch}Wt\,\,,
\end{equation}
	\begin{equation}\label{77.3}
  \Lambda_{10}=\frac{v'^*_{1}}{\sqrt{1-v'^{*2}}}=\frac{\mathrm{sh}Wt-u_1\mathrm{ch}Wt}{\sqrt{1-u^2}}\,\,,
\end{equation}	
\begin{equation}\label{77.6}
  \Lambda_{30}=\frac{v'^*_{3}}{\sqrt{1-v'^{*2}}}=-\frac{u_3}{\sqrt{1-u^2}}\,\,,
\end{equation}
			\begin{equation}\label{77.92}
  \Lambda_{12}=\Lambda_{21}=\Lambda_{32}=\Lambda_{23}=0\,,
\end{equation}	
		\begin{equation}\label{77.93}
  \Lambda_{22}=1\,,
\end{equation}	
To calculate the remaining components of 4-vectors, we introduce the quantities
\begin{equation}\label{17.2}
  \Lambda'_{\alpha\gamma}=\Lambda'_{\gamma\alpha}=\delta_{\alpha\gamma}+\frac{1-\sqrt{1-v'^{*2}}}{v'^{*2}\sqrt{1-v'^{*2}}}\,\,v'^*_{\alpha}v'^*_{\gamma}\,\,,
\end{equation}	
then
\begin{equation}\label{77.5}
  \Lambda_{\alpha\beta}=a_{\gamma\beta}\Lambda'_{\gamma\alpha}\,\,.
\end{equation}	
Let's calculate the coefficients $\Lambda'_{\alpha\gamma}$ taking into account \eqref{39}-\eqref{41}. We get
\begin{equation}\label{89.7}
  \Lambda'_{11}=1+\frac{1}{\sqrt{1-u^2}}\cdot\frac{(\mathrm{sh}Wt-u_1\mathrm{ch}Wt)^2}{\mathrm{ch}Wt-u_1\mathrm{sh}Wt+\sqrt{1-u^2}}\,\,,
\end{equation}	
		\begin{equation}\label{89.8}
  \Lambda'_{33}=1+\frac{u^2_3}{\sqrt{1-u^2}}\cdot\frac{1}{\mathrm{ch}Wt-u_1\mathrm{sh}Wt+\sqrt{1-u^2}}\,\,,
\end{equation}	
		\begin{equation}\label{89.9}
  \Lambda'_{13}=-\frac{u_3}{\sqrt{1-u^2}}\cdot\frac{\mathrm{sh}Wt-u_1\mathrm{ch}Wt}{\mathrm{ch}Wt-u_1\mathrm{sh}Wt+\sqrt{1-u^2}}\,\,.
\end{equation}	

The next four equalities are not so obvious. Replacing $\cos\phi_W$ and $\sin\phi_W$ with their values from \eqref{42}, \eqref{43} and knowing the equalities \eqref{89.7}-\eqref{89.9}, we can prove that
\[\Lambda_{11}=a_{\gamma1}\Lambda'_{\gamma1}=a_{11}\Lambda'_{11}+a_{31}\Lambda'_{31}=\cos\phi_W\Lambda'_{11}+\sin\phi_W\Lambda'_{31}=\]
	\begin{equation}\label{77.7}
	=\left(1+\frac{1-\sqrt{1-u^2}}{u^2\sqrt{1-u^2}}\,u^2_1\right)\mathrm{ch}Wt-\frac{u_1}{\sqrt{1-u^2}}\,\,\mathrm{sh}Wt\,,
\end{equation}	
		\begin{equation}\label{77.8}
  \Lambda_{33}=a_{\gamma3}\Lambda'_{\gamma3}=a_{13}\Lambda'^{13}+a_{33}\Lambda'_{33}=-\sin\phi_W\Lambda'_{13}+\cos\phi_W\Lambda'_{33}=1+\frac{1-\sqrt{1-u^2}}{u^2\sqrt{1-u^2}}\,u^2_3\,,
\end{equation}	
		  \[\Lambda_{13}=a_{\gamma3}\Lambda'_{\gamma1}=a_{13}\Lambda'_{11}+a_{33}\Lambda'_{31}=-\sin\phi_W\Lambda'_{11}+\cos\phi_W\Lambda'_{31}=\]
			\begin{equation}\label{77.9}
			=\frac{1-\sqrt{1-u^2}}{u^2\sqrt{1-u^2}}\,u_1u_3\,\mathrm{ch}\,Wt-\frac{u_3}{\sqrt{1-u^2}}\,\mathrm{sh}Wt\,,
\end{equation}	
		\begin{equation}\label{77.91}
  \Lambda_{31}=a_{\gamma1}\Lambda'_{\gamma3}=a_{11}\Lambda'_{13}+a_{31}\Lambda'_{33}=\cos\phi_W\Lambda'_{13}+\sin\phi_W\Lambda'_{33}=\frac{1-\sqrt{1-u^2}}{u^2\sqrt{1-u^2}}\,u_1u_3\,.
\end{equation}	
The proof of equalities \eqref{77.7}-\eqref{77.91} consists of carefully multiplying the given equations by $(\mathrm{ch}Wt-u_1\mathrm{sh}Wt+\sqrt{1-u^2})^2$, expanding the brackets and reducing like terms.

Comparing now \eqref{31.3}, \eqref{30.5}, \eqref{161}-\eqref{171.1} with \eqref{74}-\eqref{77.93} and \eqref{77.7}-\eqref{77.91} we can see that the methods of determining a uniformly accelerated reference frame using a boost and using the solution of the equation of motion of a tetrad are equivalent to each other.

\subsubsection{Transformation into a uniformly accelerated frame of reference in explicit form}

The general Lorentz-M{\o}ller-Nelson transformation in 4-dimensional form is written as
\begin{equation}\label{17.6}
T=\Lambda_{\alpha 0}r'_{\alpha}+\int^{t}_{0}\Lambda_{00}dt,\,\,\,\,\,\,R_{\beta}=\Lambda_{\alpha\beta}r'_{\alpha}+\int^{t}_{0}\Lambda_{0\beta}dt.
\end{equation}	
Substituting the 4-vectors calculated in the two previous paragraphs and integrating, we can write the general transformation into a uniformly accelerated frame of reference as
\begin{equation}\label{54.1}
T=\frac{\mathrm{sh}Wt-u_1\mathrm{ch}Wt}{\sqrt{1-u^2}}\left(\,\,r'_1+\frac{1}{W}\right)-\frac{u_3}{\sqrt{1-u^2}}\,\,r'_3+\frac{u_1}{W\sqrt{1-u^2}}\,\,,
\end{equation}	
\[R_1=\left[\left(1+\frac{1-\sqrt{1-u^2}}{u^2\sqrt{1-u^2}}\,u^2_1\right)\mathrm{ch}Wt-\frac{u_1}{\sqrt{1-u^2}}\,\,\mathrm{sh}Wt\right]\left(\,\,r'_1+\frac{1}{W}\right)+\frac{1-\sqrt{1-u^2}}{u^2\sqrt{1-u^2}}\,u_1u_3\,\,r'_3-\]
\begin{equation}\label{54.2}
-\frac{1}{W}\left[1+\frac{1-\sqrt{1-u^2}}{u^2\sqrt{1-u^2}}\,u^2_1\right]\,,
\end{equation}	
\begin{equation}\label{54.4}
R_2=r'_2\,,
\end{equation}
\[R_3=\frac{u_3}{\sqrt{1-u^2}}\left[\frac{1-\sqrt{1-u^2}}{u^2}\,u_1\,\mathrm{ch}\,Wt-\,\mathrm{sh}Wt\right]\left(\,\,r'_1+\frac{1}{W}\right)+\left(1+\frac{1-\sqrt{1-u^2}}{u^2\sqrt{1-u^2}}\,u^2_3\right)r'_3-\]
\begin{equation}\label{54.3}
-\frac{1-\sqrt{1-u^2}}{Wu^2\sqrt{1-u^2}}\,u_1 u_3\,.
\end{equation}	
Here the initial velocity of the system $s$ is $-\mathbf{u}$ and it is assumed that the initial orientation of $s$ coincides with the orientation of the laboratory frame of reference.

\newpage
\begin{flushleft}
{\bf{Conclusion}}
\end{flushleft}

In conclusion, we emphasize that the well-known law of hyperbolic motion of a particle under the action of a constant force is irrelevant to the problem of uniformly accelerated motion. A uniformly accelerated frame of reference moves in such a way that its frequencies of Thomas proper precession \eqref{52} and Wigner proper rotation \eqref{46} cancel each other out, and the resulting rotation of the uniformly accelerated frame $s$ is zero, as it should be.

The most general transformation to a uniformly accelerated frame of reference can be found in at least three ways: a simple Lorentz boost (Section 2), solving the equations \eqref{66} and \eqref{67} of the inverse kinematics problem (Section 4), and solving the equations of tetrad motion (Section 5). The agreement between the parameters of the uniformly accelerated reference frame obtained by solving equations \eqref{66}, \eqref{67} or equations \eqref{31.1}, \eqref{30.1} with the parameters obtained by the Lorentz boost \eqref{36}-\eqref{38}, \eqref{42}-\eqref{44} allows even a cautious researcher to conclude that the inverse kinematics equations and the tetrad's equation of motion are equivalent, verified, and fully justified.

The three-dimensional velocity vector of the origin of the reference frame, expressed in its own coordinate system, together with the direction cosine matrix and the tetrad of orthonormal 4-vectors, of course, do not exhaust the entire set of possible kinematic methods for representing the motion of a rigid reference frame. Therefore, the problem of finding equations similar to the inverse kinematics equations or the tetrad motion equations for another possible parameterization of spacetime orientation, for example, using biquaternions, is of interest. This problem is proposed for solution to all interested specialists.

\begin{flushleft}
{\bf{Gratitude}}
\end{flushleft}

The author thanks Professor N. G. Migranov for useful advice, comments and support.

\end {document}